\documentclass{azhastm}
\usepackage{graphicx}

\begin{document}

\journalinfo{2020}{97}{00}{001}[000]

\title{Contribution of binary stars to the velocity dispersion inside OB associations with {\it Gaia} DR2 data}
\author{A.~M.~Melnik\email{anna@sai.msu.ru}   and A.~K.~Dambis
\addresstext{1}{Sternberg Astronomical Institute, Lomonosov
Moscow State University} }

\shortauthor{MELNIK AND DAMBIS}

\shorttitle{CONTRIBUTION OF BINARY STARS}

\submitted{May 21, 2020; in final form, September 08, 2020}

\begin {abstract}
We estimated the contribution of binary systems to the velocity
dispersion inside OB-associations derived from {\it Gaia} DR2 proper
motions. The maximum contribution to the velocity dispersion is given
by the systems with the period of revolution  of $P=5.9$ yr whose
components  shift by a distance of about  the diameter of the system
during the base-line time of {\it Gaia} DR2 observations. We employed
two methods to study the motion of the photocenter of the binary
system: the first one uses  the total displacement  between the
initial and final visibility periods  and the second one is based on
solving a system of $n$ equations defining the displacements  at the
times $t_n$. The first and second methods yield very similar
$\sigma_{bn}$ values of 0.90 and 0.87 km s$^{-1}$, respectively.
Taking into account the fact that orbits are elliptical slightly
decreases the inferred $\sigma_{bn}$. We estimated the
eccentricity-averaged $\overline{\sigma_{bn}}$ value to be
$\overline{\sigma_{bn}}=0.81$ km s$^{-1}$ assuming that the orbital
eccentricities of massive binary systems are distributed uniformly in
the  $e \in [0,0.9]$ interval. The choice of the exponent $\gamma$ in
the power-law distribution, $p_q \sim q^\gamma$, of the
component-mass ratios $q=M_2/M_1$ of binary systems appears to have
little effect on $\sigma_{bn}$. A change of $\gamma$ from  0 (flat
distribution) to -2.0 (preponderance of systems with low-mass
components) changes  $\sigma_{bn}$ from  0.90 to 1.07 km s$^{-1}$.
\end{abstract}

\section{1. INTRODUCTION}

The second intermediate data release from the satellite {\it Gaia}
({\it Gaia} DR2) contains high-precision proper motions for 1.3
billion stars derived from the position measurements collected during
1.8 year [1], [2], [3]. The average uncertainty in proper motions for
high-luminosity stars of OB associations  in the extended solar
neighborhood is 0.1 mas yr$^{-1}$, which corresponds to the velocity
uncertainty of 0.5 km s$^{-1}$ at the distance of 1 kpc.

OB associations are sparse groups of  stars of  spectral types O and B
[4]. Blaha and Humphreys [5] compiled a catalog of high luminosity
stars in the extended solar neighborhood. Their list includes O--B2-type
main-sequence stars, O--B3-type bright giants, and supergiants of all spectral
types whose  ages do not  exceed 40 Myr. Blaha and Humphreys [5]
identified 91 OB associations located within ~3 kpc from the Sun. Of
2209 stars in OB associations 2007 (90\%) were cross-matched with
{\it Gaia} DR2 catalog.

A number of factors contribute to the calculated velocity dispersion inside OB-associations:
(1) turbulent motions inside giant molecular clouds, from which
young stars form [6]; (2) motions inside binary systems, and (3)
uncertainties of stellar velocities.

There is extensive  evidence that giant molecular clouds are  close
to virial equilibrium (e.g., [7], [8]). Stars of
OB-associations are born in the turbulent gas medium and inherit its
velocity dispersion. So the virial masses of OB associations must be
nearly equal to those of their parent molecular clouds.
The virial mass of an OB-association can be calculated by the following
formula:

\begin{equation}
 M_{vir}=\frac{5a\sigma_t^2}{G},
\label{mvir}
\end{equation}

\noindent where $a$ is the specific radius of an association and
$\sigma_t$ is one-dimensional velocity dispersion of turbulent
motions. The virial masses of OB associations from the catalog by
Blaha and Humphreys [5] lie in the interval 10$^5$--10$^7$$M_\odot$
[9], [10], which, on the whole, is consistent with  the mass estimates  of
giant molecular clouds: $10^5$--$2\times10^6$ M$_\odot$ [11].

We use {\it Gaia} DR2 proper motions to calculate the velocity
dispersions inside OB associations in the direction of the Galactic
longitude $l$ and latitude $b$. We then corrected the observed velocity dispersions,
$\sigma_{l,\,obs}$ and $\sigma_{b,\,obs}$, for the
errors in proper motions and distances:

\begin{equation}
\begin{array}{l}
\sigma_{vl}^2=\sigma_{l,\,obs}^2-(4.74\; r\, \varepsilon_{\mu
l})^2-(4.74\; a\, \overline{\mu_l})^2 \\ [10 pt]
\sigma_{vb}^2=\sigma_{b,\,obs}^2-(4.74\; r\, \varepsilon_{\mu
b})^2-(4.74\; a\, \overline{\mu_b})^2\\ [10 pt]
\end{array}
\label{smu}
\end{equation}

\noindent where $\varepsilon_{\mu l}$ and $\varepsilon_{\mu b}$ are
the average uncertainties in {\it Gaia} DR2 proper motions;
$\overline{\mu_l}$ and $\overline{\mu_b}$ are average proper motions
of stars in an association in the direction of the Galactic longitude
and latitude, respectively; $r$ is the heliocentric distance of the OB association; the
factor $4.74 \times r$ (kpc) transforms units of mas yr$^{-1}$ into
km s$^{-1}$. In our calculations of the sky-plane velocities
we use the average distance $r$ to the association and do
not use parallaxes of individual stars. In this case the average
error in distances is about the specific radius $a$ of the
association in the sky plane. The average dispersion caused by the
errors in {\it Gaia} DR2 proper motions is 0.5 km s$^{-1}$. The
errors in distances to individual stars (the last term in
Eq.~~\ref{smu}) also create an additional velocity dispersion with
the  average value equal to 0.5 km s$^{-1}$.

The average one-dimensional velocity dispersion calculated
for 28 OB associations including more than 20 stars with known {\it
Gaia} DR2 proper motions is $\sigma_v=4.5$ km s$^{-1}$. The velocity
dispersion inside OB-associations has not been corrected for
binary-star effect which was taken, by default, to be small [10].

However, the fraction of binary systems among  OB stars is quite
large and amounts to 30--100\% [12], [13], [14], [15]. In this
paper we estimate the contribution of binary systems to the velocity
dispersion inside OB-associations by  simulating the motion of binary
components.

\section{2. Results}

\subsection{2.1 Shift of the photocenter of the binary system between the
first and final {\it Gaia} DR2 visibility periods }

The shift of the component of the binary system during the {\it Gaia}
DR2 base-line time ($T=1.8$ year) gives rise to additional proper
motion and hence additional  velocity which increases the velocity dispersion inside
OB associations.

\begin{figure*}
\resizebox{\hsize}{!}{\includegraphics{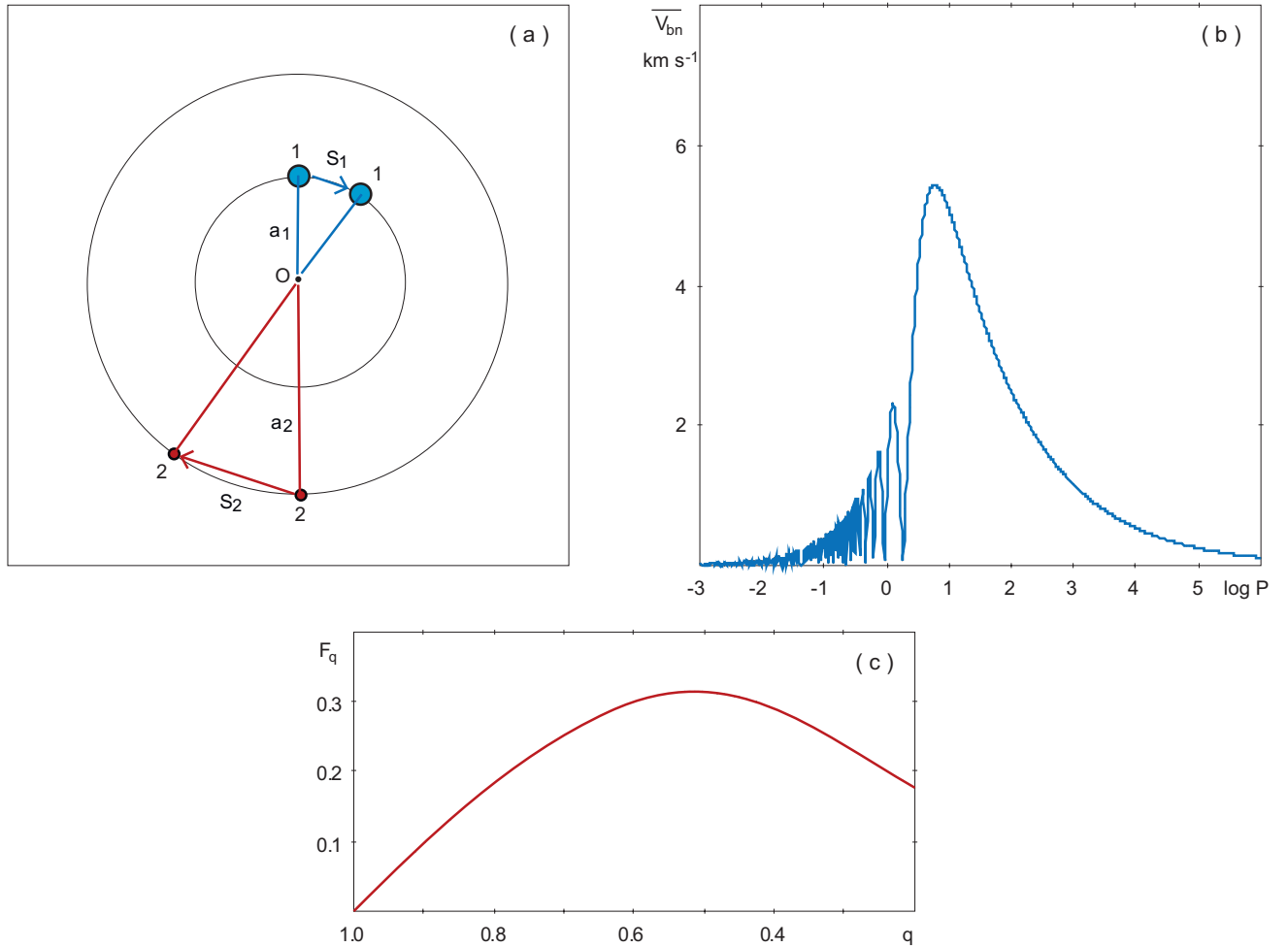}}
\caption{Contribution of binary systems to the velocity dispersion
inside OB-associations. (a) Model of a binary system with component
stars named $1$  and $2$ rotating around the common mass center in
circular orbits in the sky plane with the period of revolution $P$.
The vectors $S_1$ and $S_2$ indicate the displacements of binary
stars over the base-line time of {\it Gaia} DR2 measurements, $T$.
The mass centre is located at  point $O$. The radii of orbits of
components 1 and 2 are $a_1$ and $a_2$, respectively. (b) Dependence
of the additional velocity $\overline{V_{bn}}$ averaged over mass
ratio $q$ on $\log P$. The velocity $\overline{V_{bn}}$ is the
average contribution of a binary system with a period $P$ into the
velocity dispersion. The maximal value of the velocity
$\overline{V_{bn}}$ is 5.4 km s$^{-1}$ corresponding to the period
$P=5.9$ yr. (c) Dependence of function $F_q$ (Eq.~\ref{fq}) on mass
ratio $q$. Function $F_q$ equals zero at $q=1$ implying  the fixed
position of the photocentre of a binary system with equal-mass
components, and reaches maximum  at $q=0.52$.} \label{bin}
\end{figure*}

Let us consider a binary system with the components  moving in
circular orbits in the sky plane with the revolution period of $P$
(Fig.~\ref{bin}a). The primary and secondary components are named as
1 and 2, respectively. A typical representative of an
OB-association member star in the catalog by Blaha and Humphreys [5] is a 10
M$_\odot$ star, and we therefore assume that the mass of the primary is
$M_1=10$ M$_\odot$ and that of the secondary component is
$M_2=q\, M_1$:

\begin{equation}
q=M_2/M_1<1. \label{q}
\end{equation}

\noindent  The binary components are rotating around the mass centre
$O$ in orbits with the radii $a_1$ and $a_2$, respectively. The radius
$a_2$  can be estimated from Newton's gravity law:

\begin{equation}
a_2^3=\frac{P^2}{4\pi^2}\;\frac{G M_1}{(1+q)^2}, \label{a2}
\end{equation}

\noindent and the radius $a_1$ can be derived from the relation:

\begin{equation}
a_1 = q a_2 \label{a1}
\end{equation}

\noindent The displacements  $s_1$ and $s_2$ of binary components
between the first and final visibility periods of {\it Gaia} DR2
measurements are equal to:

\begin{equation}
S_2 =  \sqrt{2} \, a_2 \, \sqrt{1 -  \cos  \frac{2 \pi T}{P}},
\label{s2}
\end{equation}
\begin{equation}
S_1 =  q \, S_2, \label{s1}
\end{equation}

\noindent where  $T$ is the base-line time of {\it Gaia} DR2
observations.

However, {\it Gaia} DR2 observations include many visibility periods
of individual stars. For example, stars of OB-associations were
observed, on average, during $n=14$ visibility periods. The number of visibility
periods is the number of groups of observations separated from
other groups by at least 4 days [3], [16]. In section 2.2 we consider
another method based on solving a system of $n$ equations defining
the displacements $s_n$ at the times $t_n$.

Binary systems with the period of $P=5.9$ yr  appear to give the
maximal contribution to the velocity dispersion. In this case  the
displacement $S=S_1+S_2$ of the components  between the  initial and
final visibility periods of {\it Gaia} DR2 observations is nearly
equal to the diameter of the orbit $D\approx a_1+a_2$ and amounts to
$D\approx8$ a.~u.,  which corresponds to the angular separation
between the components of $\sim 8$ mas at the distance of $r=1$ kpc.
Note that the median heliocentric distance of  OB associations  in the catalog
by Blaha and Humphreys [5] is $r=1.7$, i.~e. the median
angle between the binary components must be 1.7 times smaller.

Stellar images in the {\it Gaia} focal plane have  large sizes and this
is done intentionally to increase the number of pixels
involved in image building and allow the position of the image center
to be determined with greater accuracy. The median size
of {\it Gaia} images (the full width at half maximum of the line spread
function) is $\sim 100$ mas [17]. It means that the binary stars
considered must be represented by the same source.

Thus, we should consider the displacement of the photocentre of the
binary system but not the motion of one of its components. Generally,
calculating the  effect of binaries requires a consideration of {\it
Gaia's} line spread function, point spread function and source
detection algorithm. Here we do not claim to derive the full solution
but  at the first approximation, we can suppose that the line spread
and point spread functions are nearly flat within $\pm10$ mas from
the maximum [17].

We can further assume that the light flux from both binary components
determines the position of the centre of the image. We then use the
following relation between the luminosity $L$ of a star and its mass
$M$, for example [18]:

\begin{equation}
L \sim M^4,
\end{equation}

\noindent to write the following formula for the displacement of the
photocentre of the binary system:

\begin{equation}
S_{ph}=\frac{S_1\,M_1^4-S_2\,M_2^4}{M_1^4+M_2^4}. \label{sph}
\end{equation}

\noindent   Substituting the  variables  from
Eqs~\ref{a2},~\ref{a1},~\ref{s2} and \ref{s1} we have

\begin{equation}
S_{ph}=\sqrt{2} \, \left(\frac{GM_1P^2}{4\pi^2}\right)^{1/3}
\,\sqrt{1 - \cos\frac{2 \pi T}{P}} \;\; F_q, \label{sph1}
\end{equation}

\noindent where function $F_q$ is equal to

\begin{equation}
F_q=\frac{q-q^4}{(1+q^4)\,(1+q)^{2/3}}.\label{fq}
\end{equation}

\noindent Figure ~\ref{bin}(c) shows  the variation of function $F_q$
which equals   zero at $q=1$. The motion of two identical
sources in the sky plane indeed does not produce a shift of their common
photocentre. On the other hand, if one of the components is too small
then the displacements of the primary component would be negligible.
Function $F_q$  reaches maximum at $q=0.52$.

The motion of the photocentre of the binary system creates the
additional velocity, which affects the observed stellar velocity in the sky
plane:

\begin{equation}
 V_{bn}=S_{ph}/T, \label{vbn}
\end{equation}

\noindent where $T$ is time base-line of {\it Gaia} DR2 observations.
Figure~\ref{bin}(b) shows the dependence of the additional velocity
$\overline{V_{bn}}$ averaged over mass ratio, $q$, on $\log P$. On
the whole, the dependence of $\overline{V_{bn}}$ on the period $P$
can be  easily understood: at small periods the binary systems have
small diameters and the  displacements of the photocenter are small,
the larger $P$ the larger the diameter $D$ of the system, but in this
case the orbital velocities are low and the displacements of the
photocenter   during the {\it Gaia} DR2 base-line time are small.
Consequently, there is a certain period $P$ corresponding to the
maximal shift $S_{ph}$. The maximal shift $S_{ph}$ and consequently
the maximal  velocity $\overline{V_{bn}}$ are achieved for $P=5.9$
yr. The maximal velocity is equal to 5.4 km s$^{-1}$. At small
periods the velocity $\overline{V_{bn}}$ demonstrates oscillations
and zero values $\overline{V_{bn}}=0$ correspond to the periods $P =
T/n$, where $n$ is integer and the binary system makes a whole number
of revolutions within time interval $T$, resulting in zero
displacement $S_{ph}$.

To estimate the effective contribution of binary stars,
$\sigma_{bn}$, to the velocity dispersion inside an OB-association,
$\sigma_v$, we integrate $V_{bn}^2(P,q)$ over all possible periods
and mass ratios: \vspace{4 cm}
$$
\sigma_{bn}^2= \hspace{7 cm}
$$
\begin{equation}
 f_b  f_j \int_{q=0.2}^{q=1.0}\int_{\log P=-3}^{\log
P=+6} V_b^2(P,q) f_p (P)   p(q)  d(\log P) d q, \label{sigma_b}
\end{equation}

\noindent where factors $f_b$ and $f_j$  characterize   the fraction
of binary stars and   projection effects, while functions $f_p(\log
P)$ and $f_q(q)$ take into account the period and mass-ratio distributions,
respectively. Here we make the following assumptions. The fraction of
binary systems in OB-associations amounts to $f_b=0.5$ [13]. The
ensemble of binary systems is thought to be oriented randomly with
respect to the line of sight and consequently the squared
projection of the total velocity $V$ into a random direction, for
example, $x$, amounts, on average, to $V_x^2=1/3\, V^2$, which gives
a projection factor of  $f_j=1/3$. Aldoretta et al. [19] showed
that the period distribution of massive binary stars is approximately
flat in increments of $\log P$ (so-called \"Opik's law [20],[21]) in
the interval of $\log P$ from -3 to +6. So we can suppose that
$f_p=1/9$ for unit of $\log P$. Sana and Evans [22] found that the
distribution of binary systems over mass ratio $q$ is uniform within the
interval $q=0.2$--1.0 so we can adopt $f_q=1.25$.

We now integrate  Eq.~\ref{sigma_b} numerically to determine the average
contribution of binary systems to the velocity
dispersion inside OB-associations:

\begin{equation}
\sigma_{bn}=0.90 \; (M_1/10 M_{\odot})^{1/3},
\label{sigma_b1}\end{equation}

\noindent which is more than twice  greater than $\sigma_{bn}=0.38$
km s$^{-1}$ derived for the time base-line of {\it Gaia} DR1
observations equal to $T=24$ yr [23]. Both these values of
$\sigma_{bn}$ are rather small compared to the observed velocity
dispersion, $\sigma_v\sim 4$--5 km s$^{-1}$. Eq.~\ref{sigma_b1} shows
that  $\sigma_{bn}$ depends only slightly on mass $M_1$:  a 10 times
decrease of the mass  (from $M_1=10$ to 1~M$_\odot$) translates in
only a $\sim2$ times decrease in $\sigma_{bn}$.

\subsection{2.2  Shift of the photocenter of the binary system estimated taking into account intermediate {\it Gaia} DR2 observations}

\begin{figure*}
\resizebox{\hsize}{!}{\includegraphics{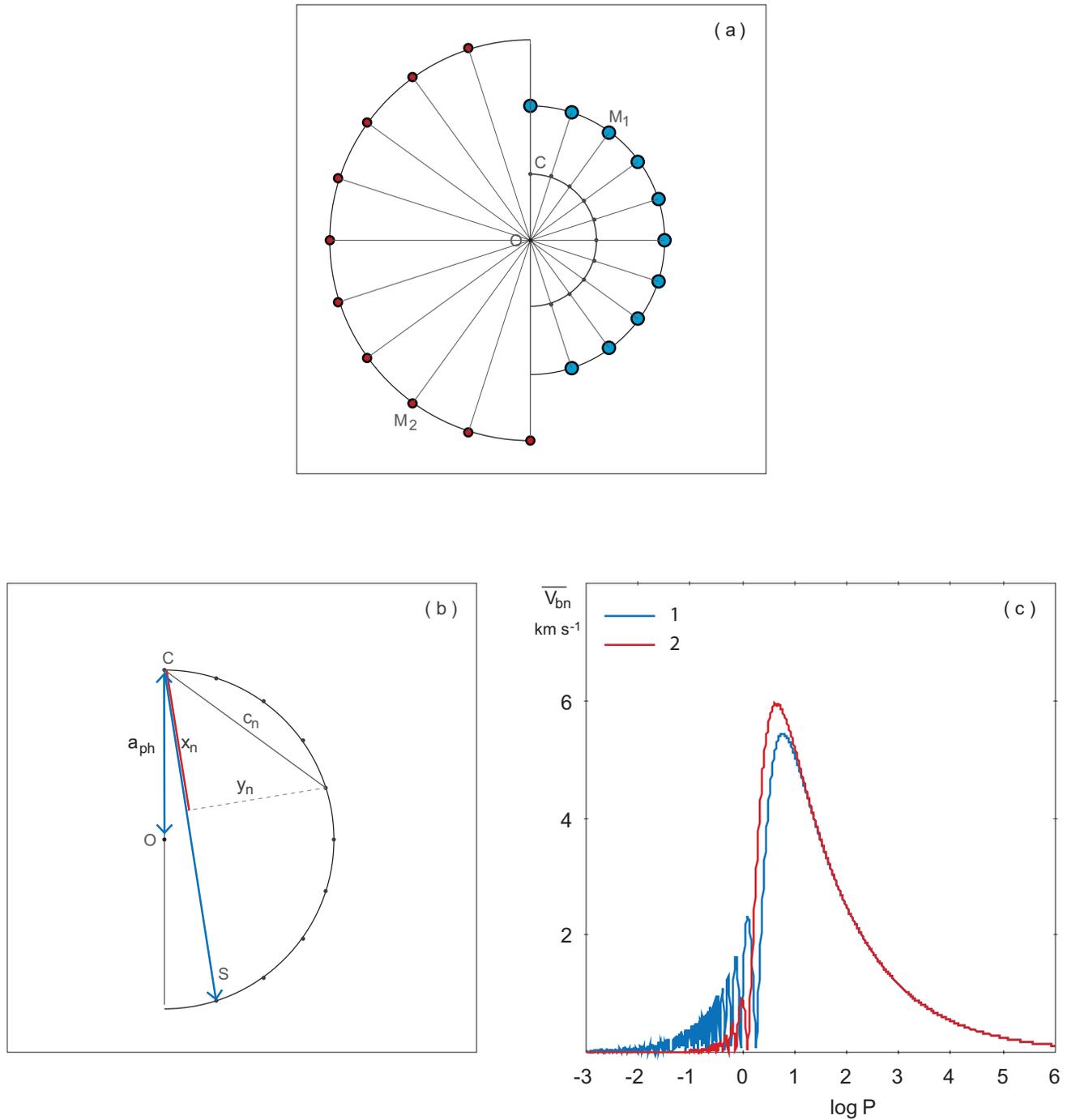}} \caption{(a) Binary
system with the components $M_1$ and $M_2$ rotating around their
common mass center  $O$. The position of the photocenter at the
initial moment is indicated by the point  $C$. (b) The motion of the
photocenter $C$ along the circle of radius  $a_{ph}$. The position of
the photocenter is measured  $n$ times during the {\it Gaia} DR2
base-line time. The vector  $S$ connects the positions of the
photocenter between the initial and final visibility periods. The
chord $c_n$ shows the displacement of the photocenter by the time
$t_n$ while $x_n$ and $y_n$  are its projections onto the direction
$S$ and onto the perpendicular direction, respectively. (c) The
dependence of the additional velocity $\overline{V_{bn}}$ averaged
over  $q$ on  $\log P$ derived by two methods: one using the total
displacement $S$ between the initial and final visibility periods
(the blue line) and another based on solving a system of $n$
equations defining the displacements $x_n$ at the times $t_n$ (the
red line). The maximal velocities $\overline{V_{bn}}$ calculated by
the first and second methods are 5.4 and 6.0 km s$^{-1}$,
respectively.} \label{bin2}
\end{figure*}

Let us assume that the position of the photocenter of the binary
system is measured $n$ times during {\it Gaia} DR2 base-line time.
For definiteness, we adopted  $n=14$, which corresponds to the
average number of {\it Gaia} DR2 visibility periods for high-luminosity
stars from the catalog by Blaha and Humphreys [5]. Figure~\ref{bin2}(a)
shows the rotation of components of the binary system
with the period of revolution equal to $P=5.9$ year providing the
maximal displacement of the protocenter during {\it Gaia} DR2
base-line time. Figure~\ref{bin2}(b) shows the rotation of the
photocenter $C$ around the mass center $O$ of the system. The radius
of the orbit of the photocenter equals:

\begin{equation}
a_{ph}= \, \left(\frac{GM_1P^2}{4\pi^2}\right)^{1/3} \;\; F_q,
\label{aph}
\end{equation}

\noindent The vector $\overrightarrow{S}$ is the displacement between
the initial and final visibility periods. Let us set the coordinate
system with the $x$ axis parallel to the vector $\overrightarrow{S}$
and the $y$ axis perpendicular to it. The chord $c_n$ shows the shift
of the protocenter by the time $t_n$. Projections of the chord
$c_n$ onto the directions $x$ and $y$ are equal to $x_n$ and $y_n$,
respectively. The maximal shift of the photocenter occurs along the
$x$ axis, but in the perpendicular direction the protocenter deviates
by a distance of $\sim|\overrightarrow{S}|/2$ and returns to close to the
initial position. The additional velocities $V_x$
and $V_y$ in the directions $x$ and $y$ can be derived by solving the
following systems of equations:

\begin{equation}
\begin{array}{l}
x_n= x_0+ V_x \,t_n,\\ [10 pt]
y_n= y_0+ V_y \,t_n.\\ [10 pt]
\end{array}
\label{xnyn}
\end{equation}

\noindent respectively. Since we calculate one-dimensional velocity
dispersion $\sigma_{bn}$ and take projection
effect into account through  the factor $f_j=1/3$, we have to find the
maximal possible displacement in one direction. It means that:

\begin{equation}
V_{bn}= V_x. \label{vbn1}
\end{equation}

\noindent Note that the velocity $V_y$ and the additional proper
motion (Eq.~\ref{mu_y}) are calculated with a large error, which could
lead to the  large size of error ellipsoid   in  determination of 5
astrometric parameters ($\alpha$, $\delta$, $\mu_\alpha$,
$\mu_\delta$ and $\varpi$) of  a star and causes their subsequent
rejection [3]. This problem will be discussed in section 2.5.

Figure~\ref{bin2}(c) shows the dependence of the additional velocity
$\overline{V_{bn}}$ averaged over $q$ on $\log P$ based on two
methods: one using the total displacement $S$ between the initial and
final visibility periods (section 2.1) and another based on solving a
system of $n$ equations defining the displacements $x_n$ at the times
$t_n$. The maximal velocities $\overline{V_{bn}}$ calculated by the
first and second methods are 5.4 and 6.0 km s$^{-1}$, respectively.
The maximal velocity $\overline{V_{bn}}$ calculated by the second
method has a larger value ($\overline{V_{bn}}=6$ km s$^{-1}$), but at
small periods, where the oscillations occur, $\overline{V_{bn}}$
takes smaller values than the velocity calculated by the first
method.

We integrate the velocity $\overline{V_{bn}}$ over the increment $\log
P$ to obtain the average contribution of binary systems to the
velocity dispersion computed taking into account intermediate
measurements:

\begin{equation}
\sigma_{bn}=0.87 \; (M_1/10 M_{\odot})^{1/3},
\label{sigma_bn}\end{equation}

Thus, the average contributions of binary systems to the velocity
dispersion inside OB associations calculated by the two methods
differ only by  3\%.

\subsection{2.3 Elliptical orbits}

The distribution of massive binary systems  over the eccentricity $e$
depends on the period: long-period systems ($P>10$ yr) have
practically uniform distribution over the eccentricity while
short-period systems ($P<100$ days) demonstrate the excess of
circular orbits [13], [15], [24], [25].

Let us consider a binary system with the eccentricity of $e=0.5$
and estimate the contribution of these systems into the velocity
dispersion inside OB associations. Figure~3(a) shows the binary
system with the components moving in elliptical orbits whose foci
are located at the mass center $O$. The points $P$ and $A$ indicate
the positions of the pericenter and apocenter, respectively, of the orbit of the
secondary component; the angles $\theta_1$ and $\theta_2$ are counted
from the pericenter $P$ and correspond to the positions of the
secondary component at the initial $t_1$ and final $t_2$ visibility
periods of {\it Gaia} DR2 observations, respectively.

The motion of material body in an elliptical orbit is defined by
Kepler's law:

\begin{equation}
E- \sin E = M, \label{kepler}
\end{equation}

\noindent where E is eccentric anomaly and  $M$ is mean anomaly:

\begin{equation}
M=2\pi/P \, (t-t_0),
\end{equation}

\noindent where $t_0$ is the pericenter passage time. If we know the
eccentric anomaly $E$ we can calculate the true anomaly, the angle
$\theta$ between the direction to the secondary component at the
moment $t$ and the direction to the pericenter [26]:

$$
 \sin \theta=\frac{\sqrt{1-e^2}\, \sin E}{1-e \cos E},
$$
\begin{equation}
\label{sin}
\end{equation}
$$
 \cos \theta=\frac{\cos E-e}{1-e \cos E }.\vspace {0.5 cm}
$$
\noindent The distance $r$ from the  mass center to the secondary
component at the time $t$ is determined by the relation:

\begin{equation}
r=\frac{a\sqrt{1-e^2}}{1+e \cos \theta }, \label{rell}
\end{equation}

\noindent where $a$ is the semi-major axis of the secondary-component
orbit, which depends on the period of the binary,  $P$, the mass of
the primary component, $M_1$,  and  the mass ratio, $q$
(Eq.~\ref{a2}).

We calculate the eccentric anomaly $E$ to an accuracy of the order of
10$^{-6}$ radian using the iterative method proposed by Danby [27].
In zero approximation $E_0=M+0.85e$ and each subsequent value is
determined by:
$$
E_{n+1}=E_n-\\
$$
\begin{equation}
 \frac{(M+e\sin E_n-E_n)^2}{E_n-2(M+e\sin
E_n)+M+e\sin(M+e\sin E_n) }.
\end{equation}

The time $t_1 \in [0, P]$  determines the angle $\theta_1$ and the
radius $r_1$ at the initial visibility period of {\it Gaia} DR2. The
time $t_2$ corresponds to the final visibility period of {\it Gaia}
DR2:

\begin{equation}
t_2=t_1 +T.
\end{equation}

\noindent For each time  $t_1$, we calculated the time $t_2$,
the angles $\theta_1$ and $\theta_2$, the distances $r_1$ and $r_2$,
as well as the displacement of the secondary component:

\begin{equation}
S_2=\sqrt{r_1^2+r_2^2-2 r_1 r_2 \cos(\theta_2-\theta_1)}.
\end{equation}

\noindent The displacement of the primary component $S_1$ is
determined by Eq.~\ref{s1}. With the knowledge of $S_1$ and $S_2$ we
can find the shift of the photocenter $S_{ph}$ (Eq.~\ref{sph1}) and
additional velocity $V_{bn}$ (Eq.~\ref{vbn}).

Figure~3(b) shows the dependence of the additional velocity
$\overline{V_{bn}}$ averaged over the parameters $q$ and $t_1$ ($t_1
\in [0, P]$) on $\log P$ calculated for orbits with the eccentricity
$e=0.5$. In that case the maximal velocity amounts to
$\overline{V_{bn}}=4.8$ km s$^{-1}$. For comparison, we also present
the velocity $\overline{V_{bn}}$ obtained for circular orbits by the
method outlined in section 2.1. As is evident from the figure, the assumption of the
orbit ellipticity slightly  decreases  the additional
velocity $\overline{V_{bn}}$.

The average contribution of elliptical orbits with $e=0.5$ to the
velocity dispersion inside OB associations is:

\begin{equation}
\sigma_{bn}=0.82 \; (M_b/10 M_{\odot})^{1/3},
\label{sigma_be}\end{equation}

\noindent which is $\sim10$\% less than the value of 0.90 km s$^{-1}$
obtained for circular orbits (Eq.~\ref{sigma_b1}). This is probably
due to the fact that the secondary component spends most of its time
near the apocenter of the orbit where it rotates with a low angular
velocity shifting by a small angle during the {\it Gaia} DR2 base-line
time, and its large distance $r$ from the mass center cannot
compensate the low angular velocity.

It turned that the change of the eccentricity  of orbits of binary
systems from $e=0$ to 0.9 results in a change in $\sigma_{bn}$ from
$\sigma_{bn}=0.90$ to 0.58 km s$^{-1}$. Assuming that the orbital
eccentricities $e$ of massive binary systems are distributed
uniformly over the interval $e \in [0,0.9]$, we calculated the  value
of $\overline{\sigma_{bn}}$ averaged over eccentricity:

\begin{equation}
\overline{\sigma_{bn}}=0.81 \; (M_b/10 M_{\odot})^{1/3},
\label{sigma_be}\end{equation}

\noindent which practically coincides with the result obtained for the
eccentricity of $e=0.5$.

\begin{figure*}
\resizebox{\hsize}{!}{\includegraphics{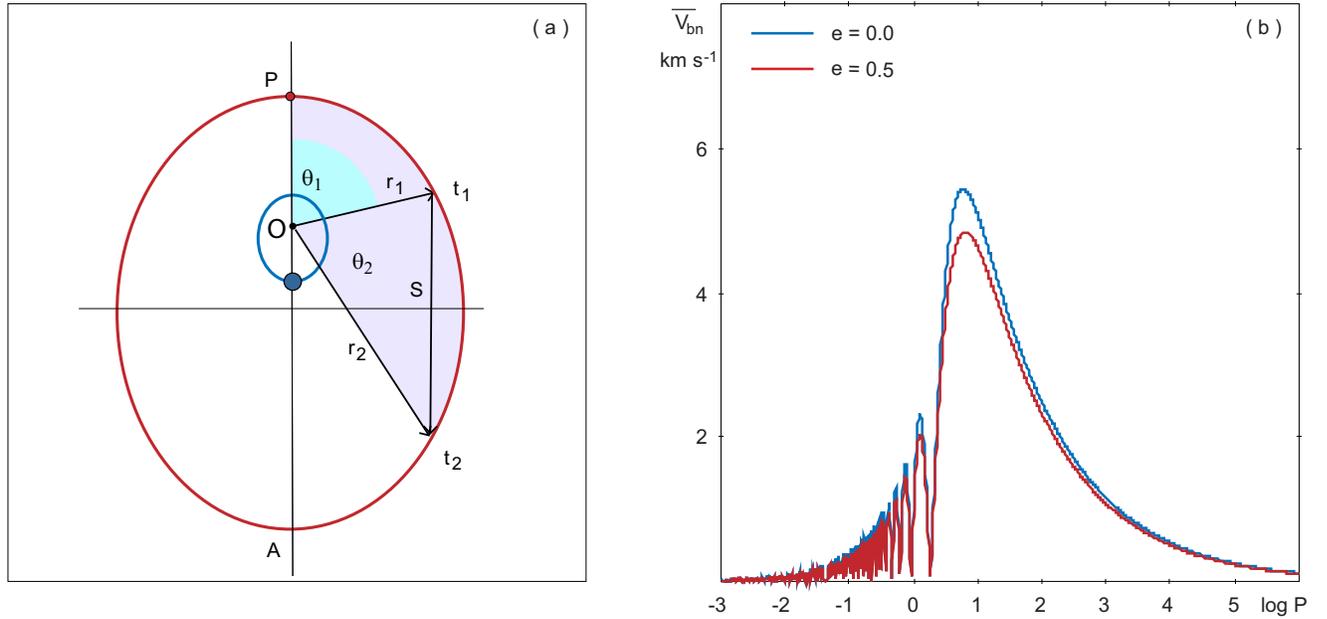}} \caption{ (a)
Elliptical orbits of components 1 and 2 of the binary system are
shown as the small and large ellipses, respectively. The foci of the
orbits and the system mass center are located at the point $O$. The
points $P$ and $A$ indicate the positions of the pericenter and
apocenter of the orbit of the secondary component. The angles
$\theta_1$ and $\theta_2$ are counted from the pericenter $P$ and
correspond to the positions of the secondary component at the initial
$t_1$ and final $t_2$ visibility periods of {\it Gaia} DR2
observations, respectively. (b) Dependence of the additional velocity
$\overline{V_{bn}}$ averaged over the parameters $q$ and $t_1$ ($t_1
\in [0, P]$) on $\log P$ calculated for orbits with the eccentricity
$e=0.5$ (the red line) and $e=0$ (the blue line).} \label{ell_1}
\end{figure*}

\subsection{2.4 Another distribution of  $p_q$}

In previous sections we suppose that the mass ratios $q=M_2/M_1$ of massive binary systems are
distributed uniformly over the interval $q
\in [0.2, 1]$ [20]. However, there is an opinion that the
distribution of the mass ratios of binary stars $p_q$ is shifted towards small
$q$  and can be described by a power law $p_q \sim q^\gamma$, where
$\gamma$ lies in the range from -0.5 to -2.4 [15].

\begin{figure*}
\centering  \resizebox{8 cm}{!}{\includegraphics{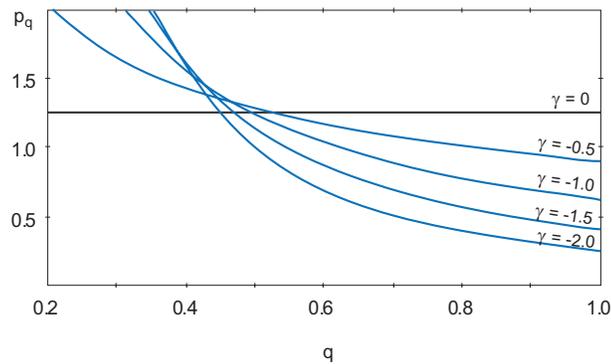}}
\caption{Distribution of component mass ratios $q$ of binary systems:
$p_q=C\,q^{\gamma}$, for $\gamma=0$ (flat distribution),
$\gamma=-0.5$, -1.0, -1.5, and -2.0.} \label{pq}
\end{figure*}

Figure~\ref{pq} shows the probability distribution
$p_q=C\,q^{\gamma}$ obtained for $\gamma=0$ (constant distribution),
$\gamma=-0.5$, -1.0, -1.5, and -2.0. The  constant $C$ is derived
from the normalization condition:

\begin{equation}
\int_{q=0.2}^{q=1.0} p_q=1. \label{cpq}
\end{equation}

\noindent Integrating $V_{bn}^2$ (Eq.~\ref{sigma_b}) with the
function $p_q=C\,q^{\gamma}$ for  exponent $\gamma=-1.5$ yields the following
velocity dispersion $\sigma_{bn}$:

\begin{equation}
\overline{\sigma_{bn}}=1.07 \; (M_b/10 M_{\odot})^{1/3},
\label{sigma_bq}\end{equation}

\noindent which is greater than  $\sigma_{bn}=0.90$
km s$^{-1}$ calculated for the flat distribution $p_q$ by only 18\%. On the
whole, the choice of exponent $\gamma$ has little effect on the the inferred
$\sigma_{bn}$: a change of $\gamma$ from  0 to -2.0 results in a variation  $\sigma_{bn}$ in the range
from 0.90 to 1.07. This is due
to the fact that the  maximal contribution to the velocity dispersion
is given by systems with the component mass ratio $q=0.5$ at which all functions $p_q$
considered have close values (Fig.~\ref{pq}).

\subsection{2.5 Analysis of errors in proper motions caused by motions of  the photocenter of  binary system}

The maximum contribution to the velocity dispersion derived from {\it
Gaia} DR2 data is provided by binary systems with the period of $P=5.9$
year. However, the photocenter moves practically linearly in one
direction and makes a swing in the other direction, which has nothing
in common with linear motion. In this section we analyze  errors in
proper motions due to nonlinear motions.

\begin{figure*}
\resizebox{\hsize}{!}{\includegraphics{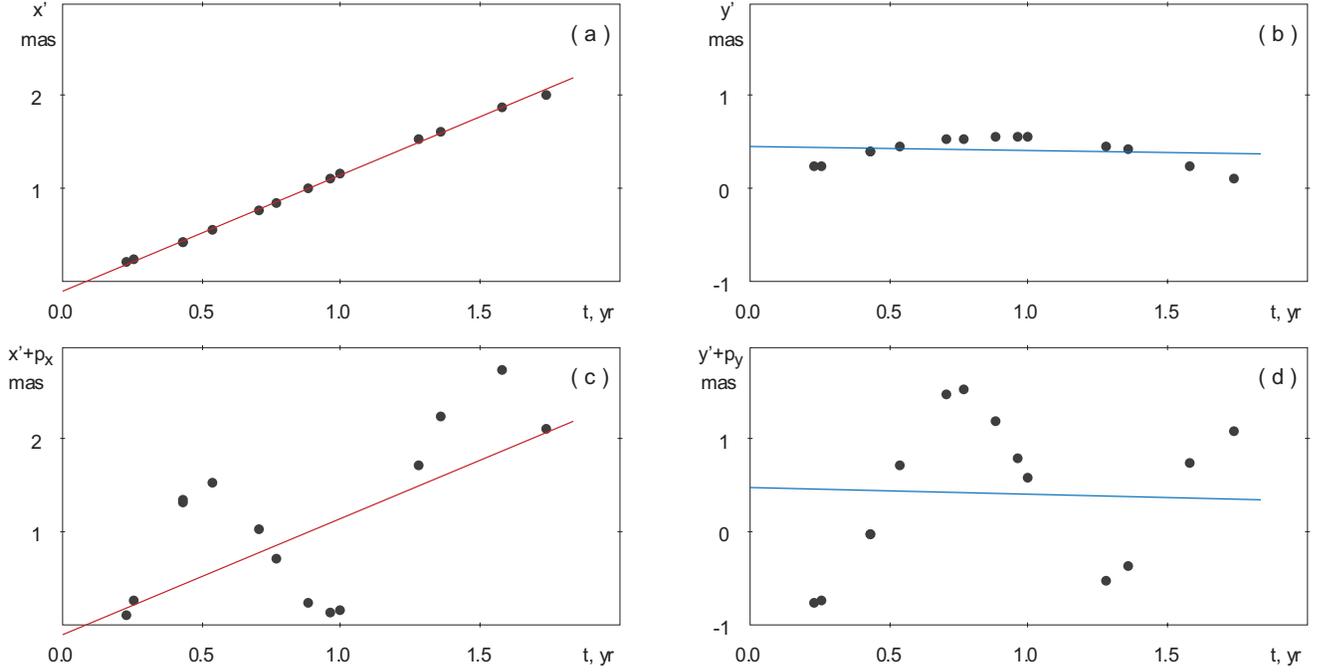}} \caption{Positions
of the photocenter of the binary system at different time instants
(points) and the proper motions derived from them (straight lines).
The $x$ axis coincides with the vector  $S$ connecting the positions
of the photocenter between the initial and final visibility periods
and the  $y$ axis is perpendicular to the $x$ axis
(Fig.~\ref{bin2}b). As an  example we chose a binary system with the
period $P=5.9$ yr providing the maximal contribution to the shift of
the photocenter. For convenience of the transformation into angular
values, the binary system was placed at the distance of $r=1$ kpc
from the Sun. (a, b) Positions of the photocenter obtained without
taking into account  the parallactic displacement. (c, d) positions
with the parallactic displacement taken into account. We can see that
the slopes of the straight lines in panels (a) and (c) as well as in
panels (b) and (d) are almost the same and this fact indicates that
the parallactic displacement has only a slight effect on the derived
proper motions. The displacement of the photocenter along the$x$ axis
(a) obeys a linear law while the displacement along $y$ axis (b)
outlines an arc, resulting in a significant increase of the errors in
the determination of the coordinate $y_0$ and proper motion $\mu_y$.}
\label{bin3}
\end{figure*}

Figure~\ref{bin3} shows the positions of the  photocenter of a
binary system at different times, and the proper motions
derived from these positions. The times are chosen randomly within the
base-line time interval of {\it Gaia} DR2. The $x$ axis coincides with the
vector $\overrightarrow{S}$ connecting the positions of the
photocenter between the initial and final visibility periods and the
$y$ axis is perpendicular to the $x$ one (Fig.~\ref{bin2}b). We take as an example
a binary system with the period $P=5.9$, which gives the
maximal contribution to the shift of the photocenter. For convenience
of the transformation into angular units, we place the binary system at a heliocentric
distance of $r=1$ kpc. The proper motions
of the photocenter, $\mu_x$ and $\mu_y$, are derived by solving the
system of linear equations defining the sky positions of the photocenter
($x_n'$, $y_n'$) at different times.

\begin{equation}
\begin{array}{l}
x_n'= x_0'+ \mu_x \,t_n,  \\ [10pt]
 y_n'= y_0'+ \mu_y \,t_n. \\ [10pt]
\end{array}
\label{xn'yn'}
\end{equation}

\noindent where  $x_n'$, $x_0'$, $y_n'$ and $y_0'$ are in the units
of mas and the  proper motions, $\mu_x$ and $\mu_y$,  are in the
units of mas yr$^{-1}$:

\begin{equation}
\begin{array}{l}
x_n'= 2.06265 \; 10^8 \;x_n/r, \\ [10pt]

x_0'=2.06265\;10^8\;x_0/r, \\ [10pt]

y_n'= 2.06265 \; 10^8 \;y_n/r, \\ [10pt]

y_0'= 2.06265 \; 10^8\;y_0/r, \\ [10pt]
\end{array}
\label{y0}\end{equation}

\begin{equation}
\begin{array}{l}
\mu_x= V_x/(4.74\,r), \\[10pt]

\mu_y= V_y/(4.74\,r), \\[10pt]
\end{array}
\label{mu_y}
\end{equation}

\noindent Figures~\ref{bin3}(a) and (b) show that  the displacement
of the photocenter along the $x$ axis obeys a linear law while the
displacement along the $y$ axis outlines an arc, significantly increasing
the errors in  determination of the coordinate $y_0$ and proper
motion $\mu_y$. On the whole, the motion along an arc implies at
least quadratic dependence on time.

We also consider the influence of the parallactic displacement on the
derived values of proper motions. Figures~\ref{bin3}(c) and (d) show
the motion of the photocenter of the binary system distorted by the
parallactic displacement. For convenience, the amplitudes of
parallactic motions along the $x$ and $y$ axes were adopted to be
$P_x=P_y=1$ mas.

\begin{equation}
\begin{array}{l}
p_x= P_x \, \cos(2\pi t_n +\varphi_0) \\ [10 pt]

p_y= P_y \,\sin(2\pi t_n +\varphi_0) \\ [10pt]
\end{array}
\label{px}
\end{equation}

\noindent We can see that the slopes of straight lines in
Figs.~\ref{bin3}(a) and \ref{bin3}(c) as well as in Figs.~\ref{bin3}(b)
and \ref{bin3}(d) are almost the same, indicating that the parallactic displacement
has only a minor effect on the derived values of
proper motions.

Table~1 lists proper motions, $\mu_x$ and $\mu_y$, and their errors
as well as the errors of the coordinates $x_0$ and $y_0$ (the
coordinates themselves are not of interest in our case) calculated
for the motion of the photocenter of the binary system with and
without taking into account the parallactic displacement. We can see that the
accuracy of the determination of the coordinate and proper motion
along the $x$ axis is almost one order of magnitude higher than along the $y$ axis. For
example, the errors in the determination of $x_0$ and $\mu_x$ are
0.01 mas and 0.01 mas yr$^{-1}$, respectively, but the errors in $y_0$ and $\mu_y$
amount to 0.1 mas and 0.1 mas yr$^{-1}$, respectively.

Thus, the motion of the photocenter of the binary system practically
does not increase  the uncertainty in the determination of   the proper
motion in one direction and adds an extra uncertainty of 0.1 mas
yr$^{-1}$ in the  determination of  the proper motion in the other
direction. The average uncertainty of proper motions of stars of OB
associations is 0.1  mas yr$^{-1}$ so the binary system considered
will have the uncertainties in the proper motions in the range
0.1--0.2 mas yr$^{-1}$.

Lindegren et al. [3] formulated three conditions under which  a
five-parameter solution derived from {\it Gaia} DR2 measurements
would not be rejected:

\begin{equation}
\left.
\begin{array}{ll}
 \verb"(1)" & \verb"mean magnitude G" \le  21.0^m\\
 \verb"(2)" & \verb"visibility_periods_used" \ge 6 \\
 \verb"(3)" & \verb"astrometric_sigma5d_max" \le \verb"(1.2 mas)" \times  \gamma(G)\\
 \end{array}
  \right \}   \\
 \label{cond}
\end{equation}

\noindent where $\gamma(G)=\max[1, 10^{0.2(G-18)}]$.

The average $G$-band magnitude and the average number of visibility
periods for member stars of OB-associatiobs are $\overline{G}=8.5^m$
and $n_\textrm{vis}=14$, respectively. Hence the first two conditions
are easily satisfied. The third condition is that a five-dimensional
equivalent to the semi-major axis of the position error ellipse,
\verb"astrometric_sigma5d_max", must not exceed 1.2 mas [3]. The
nonlinear motions of  the photocenter of the binary system considered
produce the errors at the level of 0.1 mas, which is nearly an order
of magnitude less than the threshold value of 1.2 mas
(Eq.~\ref{cond}). Thus, the calculated five-parameter solution for
the motion of the photocenter of the binary system will not be
rejected.

Note that the binary systems of other periods produce smaller
displacements of the photocenter and, consequently, the errors due to
nonlinear motions must be smaller as well.

Furthermore, the median heliocentric distance of OB associations  in the
catalog by Blaha and Humphreys [5] is $r=1.7$ kpc and, consequently,
the median errors of the proper motions must be 1.7
times smaller.

\begin{table*}
\centering \caption{Errors in coordinates and proper motions}
 \begin{tabular}{lcccc}
 \\[-7pt] \hline\\[-7pt]
& $\pm\varepsilon_{x0}$ & $\pm\varepsilon_{y0}$ & $\mu_x\pm\varepsilon_{\mu x}$ & $\mu_y\pm\varepsilon_{\mu y}$   \\
& mas & mas & mas yr$^{-1}$ &  mas yr$^{-1}$ \\
 \\[-7pt] \hline\\[-7pt]
$P_x=P_y=0$ & $\pm0.012$ & $\pm0.085$ & $1.248\pm0.012$ &
$-0.046\pm0.086$ \\
 \\[-7pt] \hline\\[-7pt]
$P_x=P_y=1$ mas & $\pm0.012$ & $\pm0.085$  & $1.260\pm0.009$ &
$-0.077\pm0.090$ \\
 \\[-7pt] \hline\\[-7pt]
\end{tabular}
\end{table*}

\section{3. Conclusions}

We estimated the contribution of binary systems to the velocity
dispersion inside OB-associations derived from {\it Gaia} DR2 proper
motions. Its average value is $\sigma_{bn}=0.90$ km s$^{-1}$. The
maximum contribution to the velocity dispersion is provided by the
systems with the period of revolution  of $P=5.9$ yr whose components
shift at the distance close to the diameter of the system during the
base-line time of {\it Gaia} DR2 observations. These systems have a
diameter of $\sim 8$ a.~u., which corresponds to the angular
separation of $\sim 8$ mas at the distance of 1 kpc from the Sun. The
characteristic size of images in the focal plane of {\it Gaia}
satellite is $\sim 100$ mas, implying that both components of the
binary system form a single image.

We studied the motion of the photocenter of the binary system by two
methods: one using the total displacement $S$ between the initial and
final visibility periods  and another based on solving a system of
$n$ equations defining the displacements $x_n$ at the times $t_n$. As
stars of OB-associations [5] were observed, on average, during 14
visibility periods, we adopted $n=14$. The first and second methods
yield very similar values of 0.90 and 0.87 km s$^{-1}$.

We analyzed the effect of  ellipticity of orbits  of binary stars
on the estimated velocity dispersion inside OB associations $\sigma_{bn}$. It
turns that accounting for the orbit eccentricity decreases
$\sigma_{bn}$ by 10\%. For example, orbits with the  eccentricity
of $e=0.5$ yield  $\sigma_{bn}=0.82$ km s$^{-1}$. Assuming
that the orbital eccentricities of massive binary systems are
distributed uniformly over the interval $e \in [0,0.9]$, we obtained
the eccentricity-averaged $\overline{\sigma_{bn}}$
to be $\overline{\sigma_{bn}}=0.81$ km s$^{-1}$.

The choice of exponent $\gamma$ in the power distribution $p_q \sim
q^\gamma$, where $q=M_2/M_1$, appears to have little effect on the
inferred $\sigma_{bn}$. A change of $\gamma$ from  0 (flat
distribution) to -2.0 (preponderance of low-mass components)
results in  the variation of $\sigma_{bn}$ from  0.90 to 1.07 km
s$^{-1}$.

Binary systems increase the errors in the determination of coordinates
and proper motions by  $0.1 \times r^{-1}$ mas and $0.1
\times r^{-1}$ mas s$^{-1}$, respectively. Given that the median heliocentric
distance of OB associations is $r=1.7$ kpc the
errors resulting from  nonlinear motion of the photocenter are almost
by an order of magnitude less than  the threshold  size of the error ellipsoid equal
to 1.2 mas above which the five-parameter
solution derived from {\it Gaia} data is rejected. Thus,  the nonlinear motions
of photocenters of binary systems do not cause a crucial increase of
errors in determination of astrometric parameters.

We showed that the contribution of binary systems to the velocity
dispersion inside OB-associations ($\sigma_{bn}=0.8$--1.1 km
s$^{-1}$) is small compared to the velocity dispersion caused by
turbulent motions inside giant molecular clouds, $\sigma_t=4$--5 km
s$^{-1}$.

The allowance for the effect of binary stars decreases the estimate
of the velocity dispersion caused by turbulent motions. Its average
value calculated for 28 OB associations decreases from 4.5 to 4.4 km
s$^{-1}$, resulting in the increase of the  median star-formation
efficiency from 1.2 to 1.3\% (for more details see [10]). The
inferred low star-formation efficiency inside giant molecular clouds
is consistent with other estimates [28], [29], [30].

\acknowledgements

This work has made use of data from the European Space Agency (ESA)
mission {\it  Gaia} (https://www.cosmos.esa.int/gaia), processed by
the  {\it Gaia} Data Processing and Analysis Consortium (DPAC,
https://www.cosmos.esa.int/ web/gaia/dpac/consortium). Funding for
the DPAC has been provided by national institutions, in particular
the institutions participating in the {\it Gaia} Multilateral
Agreement. A.~D. acknowledges the support from the Russian Foundation
for Basic Research (project nos. 18-02-00890 and 19-02-00611).

\end{document}